\documentclass{article}
\usepackage[english]{babel}
\usepackage[letterpaper,top=2cm,bottom=2cm,left=3cm,right=3cm,marginparwidth=1.75cm]{geometry}
\usepackage{setspace} 

\usepackage{mathrsfs}
\usepackage{amsmath}
\usepackage{amssymb}
\usepackage{graphicx}
\usepackage{siunitx}
\usepackage{caption}
\usepackage{cancel}
\usepackage{subcaption}
\usepackage{authblk}
\usepackage{multirow}
\usepackage{rotating}
\usepackage[table,xcdraw]{xcolor}
\usepackage{placeins}
\usepackage{lipsum}
\usepackage{listings}
\usepackage[bottom]{footmisc}
\usepackage[colorlinks=true, allcolors=blue]{hyperref}

\NewDocumentCommand{\codeword}{v}{%
\texttt{\textcolor{blue}{#1}}%
}
\begin{document}

\title{The effect of chemical vapor infiltration process parameters on
flexural strength of porous $\alpha$-SiC: A numerical model}
\author{Joseph J. Marziale, Jason Sun, Eric A. Walker, Yu Chen, David Salac, James Chen\footnote{Corresponding Author; chenjm@buffalo.edu.}}
\affil{Department of Mechanical and Aerospace Engineering, The State University of New York at Buffalo, Buffalo, New York 14260, United States}
\maketitle

\begin{abstract}
The flexural strength variability of $\alpha$-$SiC$ based ceramics at elevated temperatures creates the need for an Integrated Computational Materials Engineering (ICME) framework that relates the strength of a specimen directly to its manufacturing process. To create this ICME framework a model must first be developed which establishes a relationship between the chemical vapor infiltration (CVI) process and parameters, the resulting mesoscale pores, and the overall macroscale flexural strength.
Here a nonlinear single pore model of CVI is developed used in conjunction with a four-way coupled themo-mechanical damage model. The individual components of the model are tested and a sample system under a four-point bending test is explored. Results indicate that specimens with an initial porosity greater than 30\% require temperatures below 1273 K to maintain structural integrity, while those with initial porosities less than 30\% are temperature-independent, allowing for optimization of the CVI processing time without compromising strength.

\end{abstract}
\section{Introduction}\label{Sec1}

Alpha silicon carbide, $\alpha$-$SiC$, also called $6H$-$SiC$, has inspired considerable research effort because of the weak relationship between its material properties and temperature~\cite{Munro1997}. This is a valuable attribute for any system that must endure a high-temperature environment. Examples of $SiC$-based ceramics being used for this reason include gas turbine shrouds, vanes~\cite{Eaton2002}, and blades~\cite{Lu2014}; furnace linings~\cite{Kumar2012}; nose cones, wings, cover plates for aerospace vehicles~\cite{Qinglong2021}; and heat exchangers~\cite{Fend2011, Takeuchi2010}. The general trends of the moduli representing elastic behavior and flexure response as a function of temperature are known; however, across studies, there is wide variation in what is understood to be the flexural strength, $\sigma_f$, of a sintered $\alpha$-$SiC$ for each temperature. The problem of flexural strength uncertainty in $\alpha$-$SiC$ is well documented in the literature. Early calculations of $\sigma_f$ for $\alpha$-$SiC$  across several experiments differed by approximately 15$\%$~\cite{Munro1997}. Follow-up statistical studies suggested that the flexural strength of samples manufactured by theoretically identical processes subjected to identical loading conditions is Weibull-distributed with a low modulus~\cite{Lu2002, Gong1999,Khalili1991}, implying a large distribution width and a large degree of uncertainty.

Experimental literature suggests that the amount of constituent defects of the ceramic has a commensurate effect on the variation associated with the elastic behavior and flexure response~\cite{Fernandez2003}. There are several types of defects which can occur. Unrepresented material impurities, which for $SiC$ commonly includes oxygen~\cite{Gali2002,Yang2016}, can contribute to the local concentration of stress~\cite{Guoping2007}. Variations in grain size and grain boundaries are instrumental in producing faults~\cite{Swaminathan2010}, as they act as dislocation obstacles, giving rise to the quality of brittleness, as well as providing sites for crack initiation. However the defect for which considerable research effort has been dedicated is the presence of porosity~\cite{Yang2019,Yuan2018,Pei2019,Naslain2005}. It is confirmed that pores forming on multiple different length scales can all contribute to material weakening~\cite{Nagaraja2020, Guan2021}. The void space inherent to porosity is in fact linked to damage directly in typical phase field descriptions of fracture~\cite{Abaza2022}. 

Any step of the manufacturing process used to fabricate $\alpha$-$SiC$ ceramics is at least in part responsible for creating imperfections, especially pores. Because of this, the infiltration of matrix material has emerged as a necessary component of the manufacturing process as a subsequent porosity controlling mechanism. $SiC$ matrix infiltration is done in a number of ways. Some relevant methods are reactive melt infiltration~\cite{Caccia2019}, liquid phase infiltration~\cite{Interrante1994}, and sol-gel infiltration~\cite{Qian2004}. However the most common methodology is chemical vapor infiltration, or CVI~\cite{Petrak2001,Lv2019,Barua2020,Li2005}, because it can be tightly controlled by the processing temperature and pressure, which leads to a high purity~\cite{Hinoki2003}. CVI is used for the fabrication of ceramic matrix composites~\cite{Lv2019,Hinoki2003,Langlais2018}, or CMCs, but the process can also be applied to monolithic porous materials. Gupte and Tsamopoulous described the densification of a porous $SiC$ preform through CVI~\cite{Gupte1989}. Both the $\alpha$- and $\beta$-$SiC$ polytypes are viable options for CVI, with the surface morphology of the latter analyzed by Huang et al~\cite{Huang2022}. A CFD model of single phase $SiC$ film growth over a fuel combustion nozzle in a hot wall reactor was presented by Bijjargi et al~\cite{Bijjargi2022}. The utility of thin, single phase chemical vapor deposited $SiC$ films for electronic applications is described by Xu et al~\cite{Xu2022}. Although more susceptible to fracture, monolithic $SiC$ ceramics possess a more favorable strength and resistance to oxidation than composite materials; to reduce the disadvantage of fracture susceptibility, densification through CVI is introduced, combining the advantages of both classes of materials~\cite{Gupte1989}. 
 
The CVI process starts with a ceramic substrate in an enclosed chamber. A gaseous precursor is introduced from one end of a pore. The typical precursor used is methyltrichlorosilane ($MTS$) for its ability to pervade at relatively low temperatures~\cite{Sotirchos1991,Ganz1996,Lu2009,Ravasio2013}. However, because the interfacial reaction between the $MTS$ precursor and the $SiC$ solid occurs at the entrance location, it is where the pores close first. After closure of the interface, further gaseous infiltration is prevented, leaving void space towards the top side. The result of a large number of pores is the increase of uncertainty in the material's flexure response to the loading conditions, which can be demonstrated with the failure probability equation~\cite{Zok2017}.  In addition to uncertainties, most deposition reactions utilized in CVI processing are relatively slow, often causing physical CVI experiments to take hundreds of hours \cite{Sugiyama1989, Li2021}. Also, CVI testing requires a non-negligible budget dedicated to laboratory equipment and material samples \cite{Roman1994}.  

With cost, time, and especially uncertainty acting as barriers to physical experiments, it is therefore crucial to be able to computationally  map the flexure strength and performance of $\alpha$-$SiC$ directly as a function of the process parameters that go into the porosity controlling mechanism of CVI. The need for a hierarchical relationship between processing, structure, properties, and performance was first articulated by Olson~\cite{Olson1997,Olson1998} under the name Integrated Computational Materials Engineering (ICME), and then by Hirsch~\cite{Hirsch2006} as Through Process Modelling. For the simulation of metals, research done in view of the ICME hierarchy over the past few decades has been thorough~\cite{Horstemeyer2012,Zhang2021,Wang2020,Raturi2019,Taylor2018}. And although the concept of ICME is fully transferable to ceramics, its high potential is only recently realized~\cite{Shi2012,Raether2018,Sun2022-1}. To the author's knowledge, there does not currently exist a unified computational framework that directly connects the process parameters of chemical vapor infiltration to the flexure response of $\alpha$-$SiC$. The purpose of this study is to relate the structure of a CVI-treated porous $\alpha$-$SiC$ ceramic to its flexural strength directly as a function of the process parameters of pressure, temperature, reaction kinetics, and gas diffusivity. 

The structure of the rest of this study is as follows. Section \ref{Sec2} is a description of the CVI modeling strategy and its connection to a fracture prediction model, spanning an ICME framework. Section \ref{Sec3} provides an analysis of the relationship between several CVI process parameters and the flexure response of the resulting material, as well as an interpretation of these results as they apply to a manufacturing environment. Section \ref{Sec4} concludes the study. 
\section{Integrated Computational Materials Engineering Framework}\label{Sec2}
\begin{figure}[ht]
    \centering
    \begin{subfigure}[c]{6cm}
        \includegraphics[width=\linewidth]{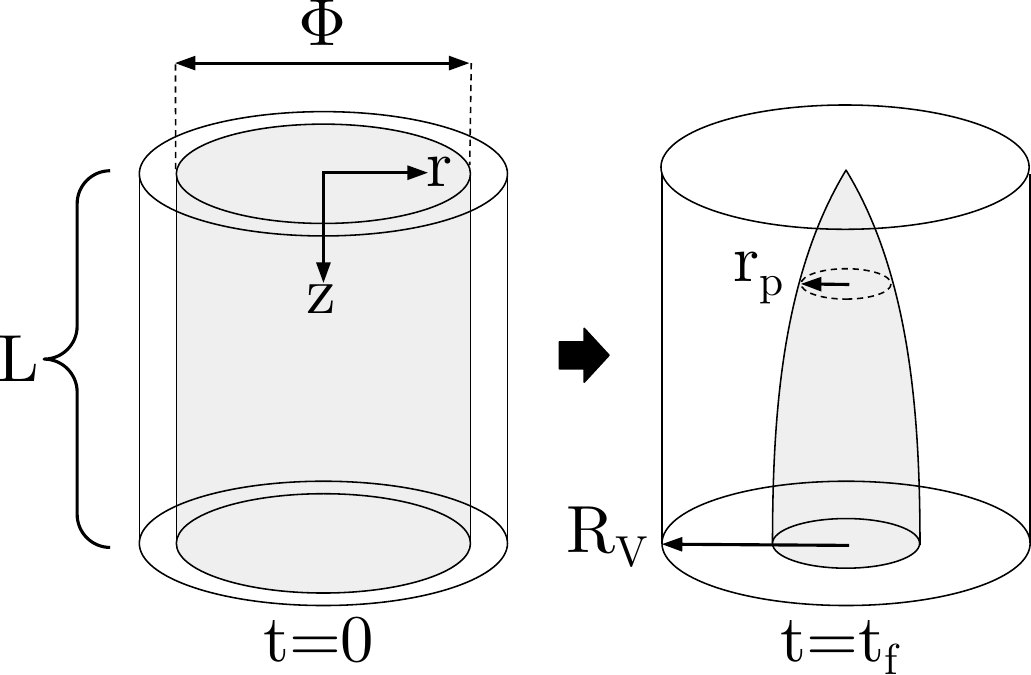}
        \caption{\doublespacing \ }
        \label{fig:CVI-FEM-1}
    \end{subfigure}\\
    \smallskip
    \begin{subfigure}[c]{10cm}
        \includegraphics[width=\linewidth]{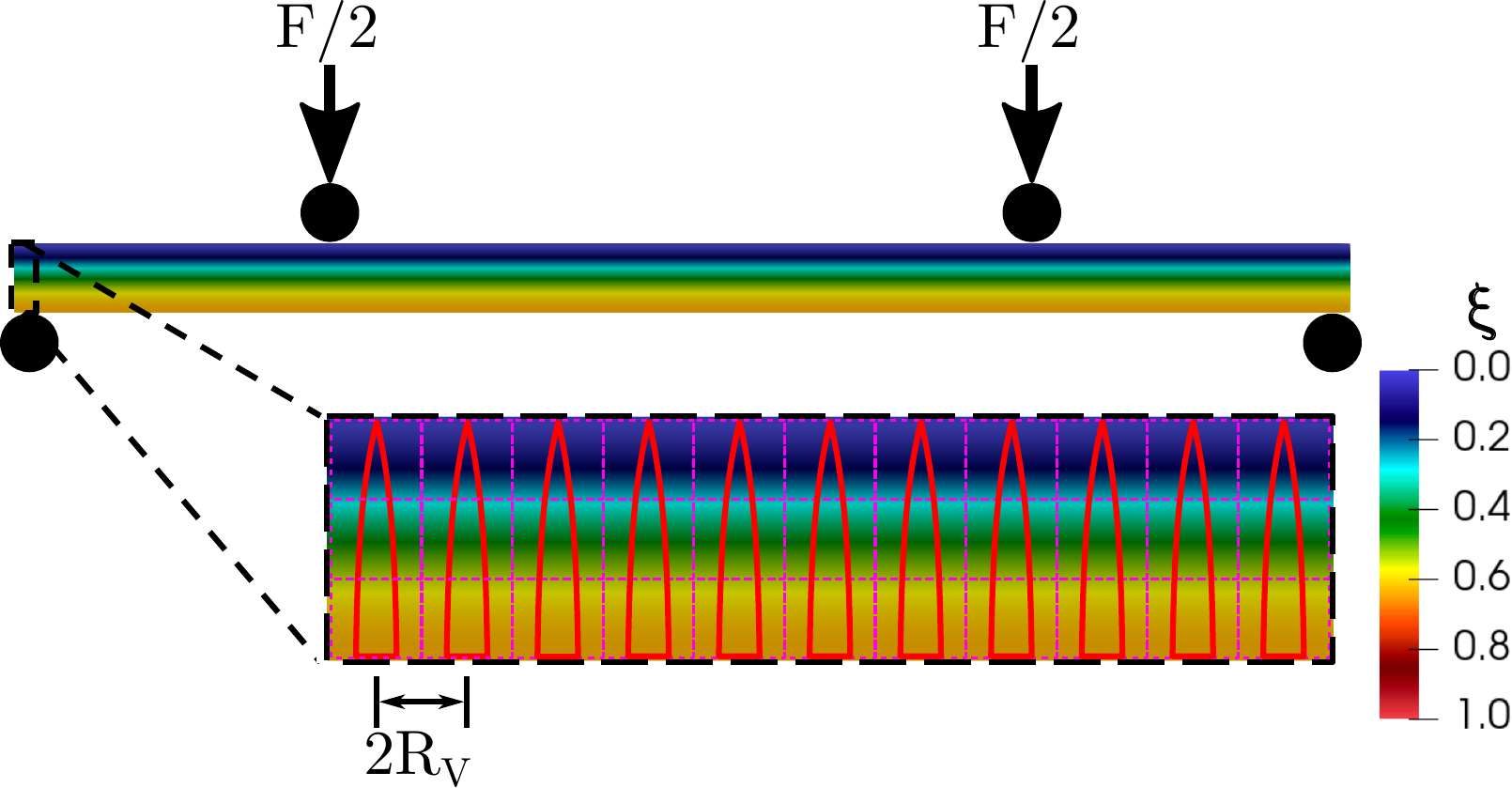}
        \caption{\doublespacing \ }
        \label{fig:CVI-FEM-2}
    \end{subfigure}
    \caption{\doublespacing
The two steps of the modeling process. (a): Evolution of a representative pore during CVI processing.  (b): Four-point bending test with representative volume elements containing the pore shape obtain from the CVI processing.}
    \label{fig-cvifs}
\end{figure}

This section describes the mathematical formulation that connects the CVI processing to the mechanical performance of a porous specimen. This is a two-step process. First, the pore shape in a representative volume element (RVE) is evolved until the entrance location closes, stopping the infiltration of the precursor, as in Fig.~\ref{fig:CVI-FEM-1}. Second, the resulting RVE is used to create a finite element model of a four-point bending test, as in Fig.~\ref{fig:CVI-FEM-2}. This is accomplished by transforming the pore shape as a function of depth into a damage parameter, $\xi$, which is evolved during the virtual testing process.

\subsection{Processing to porosity}\label{sec-processingporosity}
Using NumPy, the two broad physical phenomena being modeled in this section are gas transport and pore filling~\cite{Shinde2021}. This study uses the single pore model due to its minimal complexity, which is attractive in such contexts as uncertainty quantification, where the model must have a set of process parameters that is sufficiently small as it is sampled a large number of times with changes in its inputs~\cite{Walker2022}. The single pore model has been repeatedly validated~\cite{Langlais2018, Fedou1993, Fedou1993-2, Carolan1987, Shima2017} and is summarized in this section. The adopted model is generalizable to CVI processes other than ones involving silicon carbide, as evidenced an early model of zirconia deposition done by Carolan \cite{Carolan1987}. The single pore model is in fact universal to systems involving a single deposition reaction with first order kinetics \cite{Vignoles2015}. While this model is commonly accepted, more complex pore geometries have also been considered, and the evolution of these geometries during the CVI process differ \cite{Langhoff2008}. This was most obviously demonstrated by a study comparing the mass transport of two systems of parallel pores with differing geometries (one system being pores within a singular fiber bundle, and the other being a grid of inter-bundle pores) \cite{Kulik2004}. It is found that the mass exchange in certain pore systems homogenizes the vapor composition, which alters the final infiltration time. In addition, a pore’s tortuosity increases the path length of the transporting gas \cite{Ghanbarian2013}, which increases infiltration time and residual porosity. The current study establishes a basic ICME framework relating pore radius to flexure strength, providing the foundation for future research effort in which the performance of more complex geometries \cite{Kulik2004,Ramanuj2022,Sotirchos1991-2} can be explored. Interested readers can refer to Vignoles, who reviewed various modeling strategies of CVI~\cite{Vignoles2015}.

The chemical reaction responsible for the deposition of $SiC$ is that of the $MTS$ precursor, $CH_3SiCl_3$, which decomposes according to \cite{Walker2023}
\begin{equation*}
    CH_3SiCl_{3\text{(g)}} + H_{2\text{(g)}} \longrightarrow SiC_{\text{(s)}} + 3HCl_{\text{(g)}}  + H_{2\text{(g)}} .
\end{equation*}
The concentrations of each of the gaseous species are denoted as $\{C_{MTS},C_{H2},C_{HCl}\}=\{C_0,C_1,C_2\}=C_i$. Assuming the gas concentrations only vary in the longitudinal ($z$-)direction, the conservation equation for species $i$ can be expressed as
\begin{equation}
    \frac{\partial C_i}{\partial t} + \frac{\partial N_i}{\partial z} = R_i,
\end{equation}
where $N_i$ denotes absolute molar flux in mol m$^{-2}$s$^{-1}$, $R_i$ is an algebraic term of source or sink for species $i$, $z$ is the longitudinal spatial dimension, and $t$ is time. It is assumed that the gas reaches a quasi steady-state much faster than the evolution of the pore radius and thus the time portion may be neglected, which is supported by Gupte and Tsamopolous~\cite{Gupte1989}. This leads to
\begin{equation}
    \frac{\partial N_i}{\partial z} = R_i.
\end{equation}
The term $R_i$ can be deduced by writing the conservation equation over a very thin shell of the pore which has thickness $\delta z$, cross sectional area $\mathcal{A}$, and perimeter $\mathcal{P}$. Then
\begin{equation}
    \frac{\partial N_i}{\partial z} = \frac{s_i\mu \mathcal{P}}{\mathcal{A}},
\end{equation}
where $s_i$ is the stoichiometric coefficient of gas species $i$ and $\mu$ is the reaction rate factored by the gas concentration of the functional deposition precursor $MTS$. Particularly
\begin{equation}\label{reaction-rate}
    \mu = k_0 \exp\biggl( \frac{-E_a}{RT_{\text{CVI}}}\biggl)C_{MTS},
\end{equation}
where $E_a$ is activation energy, $R$ is the ideal gas constant, and $T_{\text{CVI}}$ is temperature~\cite{Fedou1993}. Thermodynamic constants corresponding to $MTS$ are $k_0=3.89\times 10^9$ m s$^{-1}$ and $E_a=296$ kJ mol$^{-1}$. For a cylindrical pore, $\mathcal{P}=\pi \Phi$ and $\mathcal{A} = \pi \Phi^2/4$, where $\Phi$ denotes the pore diameter. This means
\begin{equation*}
    \frac{\mathcal{P}}{\mathcal{A}} = \frac{\pi \Phi}{\pi \Phi^2 / 4} = \frac{4}{\Phi}
\end{equation*}
and thus
\begin{equation}
    \frac{\partial N_i}{\partial z} = \frac{4 s_i\mu }{\Phi }.
\end{equation}
Summing over all the gases gives
\begin{equation}
    \frac{\partial \widehat{N}}{\partial z} = \frac{4\widehat{s} \mu }{\Phi },
\end{equation}
where $\widehat{N} = \sum_i N_i$ and $\widehat{s} = \sum_i s_i$. It is assumed that the absolute molar flux includes a pure diffusion term and a positive convective term, leading to the absolute molar flux~\cite{Fedou1993-2}
\begin{equation}\label{stefan-flow}
    N_i = -D_i\frac{\partial C_i}{\partial z} + x_i \widehat{N}.
\end{equation}
Therefore, the mass transfer equation to be solved becomes
\begin{equation}\label{masstransfe}
    D_i \frac{\partial ^2 C_i}{\partial z^2} - \frac{\partial (x_i \widehat{N})}{\partial z} + \frac{4s_i\mu }{\Phi } = 0.
\end{equation}
Recall that the molar fraction $x_i$ is a function of gas concentration $C_i$:
\begin{equation}
    x_i = \frac{C_i}{\sum_l C_l} = \frac{C_i}{C_i + \sum_{l\neq i}C_l}.
\end{equation}Also
Substituting this into Eq.~\eqref{masstransfe} leads to
\begin{equation}\label{nonlinear-nondiscretized}
    \frac{\partial^2 C_i}{\partial z^2} - \frac{1}{D_i}\frac{\partial}{\partial z}\biggl( \frac{C_i}{C_i + \sum_{l\neq i}C_l}\widehat{N}\biggl) + \frac{4s_i\mu}{D_i\Phi}=0,
\end{equation}
which can be discretized on a uniform grid as
\begin{equation}\label{nonlinear-discretized}
    \frac{C_{i,j-1} - 2C_{i,j} + C_{i,j+1}}{h^2} - \frac{1}{D_{i,j}}\dfrac{1}{2h}\left(\frac{C_{i,j+1}\widehat{N}_{j+1}}{C_{i,j+1} + \sum_{l\neq i}C_{l,j+1}} - \frac{C_{i,j-1}\widehat{N}_{j-1}}{C_{i,j-1} + \sum_{l\neq i}C_{l,j-1}}\right) + \frac{4s_i \mu_j}{D_{i,j}\Phi_j}= 0,
\end{equation}
where the subscript $j$ represents a value at the location $z$, $j+1$ represents a value at the next grid point $z+\Delta z$, etc.

Diffusivity has two contributions: Fick and Knudsen diffusion. This results in
\begin{equation}
    D_{i} = \biggl(\frac{1}{D_{i,K}} + \frac{1}{D_{i,F}}\biggl)^{-1}
\end{equation}
where
\begin{equation}\label{fickknudsen}
    D_{i,F} = \biggl[\frac{1}{1-x_i}\sum_{l} \frac{x_l}{D_{b,il}}\biggl]^{-1} \textnormal{ and } D_{i,K} = d_{i,K}\Phi T_{\text{CVI}}^{1/2}.
\end{equation}
The quantity $d_{i,K}$ denotes the reduced Knudsen coefficient and $D_{b,il}$ denotes the binary Fick diffusion coefficient between species $i,l$. The latter is calculated using the Gilliland formula~\cite{Gilliland1934}, which is
\begin{equation}\label{gilliland}
    D_{b,il} = 1.360\times 10^{-3} \frac{T_{\text{CVI}}^{3/2}}{P(V_i^{1/3} + V_l^{1/3})^2}\biggl[  \frac{1}{M_i} + \frac{1}{M_l}    \biggl]^{1/2}.
\end{equation}
Molar volumes $V_i$, molar masses $M_i$, and reduced Knudsen coefficients $d_{i,k}$ are provided for each species in Table \ref{physical-constants}. Due to the nature of the molar fraction term and the diffusion coefficients, the discretized equation, Eq.~\eqref{nonlinear-discretized}, results in a non-linear system to solve.

\begin{table}
\centering
\begin{tabular}{r|lll}
                                       & $MTS$ & $H_2$ & $HCl$ \\ \hline
    $M_i$ ($10^{-3}$ kg/mol)            & 149   & 2 & 36.5 \\
    $V_i$ ($10^{-6}$ m$^3$/mol) & 122.7 & 14.3  & 25.3 \\
    $d_{i,K}$ (m/s$\sqrt{\text{K}}$)   & 3.97  & 34.30 & 8.03
    \end{tabular}
\caption{\doublespacing Molar masses, molar volumes, and reduced Knudsen coefficients of each gaseous species~\cite{Fedou1993}.}
\label{physical-constants}
\end{table}

Once the concentration of $MTS$, $C_{MTS}$, is obtained, the pore diameter can be updated. The governing equation for the evolution of the pore diameter is~\cite{Vignoles2015}
\begin{equation}\label{pore-diameter}
    \frac{\partial \Phi}{\partial t} = -2\mu V_{SiC},
\end{equation}
where the reaction rate $\mu$ is defined in Eq.~\eqref{reaction-rate} and $V_{SiC}=1.25\times 10^{-5}
 \text{m}^3\text{mol}^{-1}$ is the molar volume of $SiC$. To first-order, Eq.~\eqref{pore-diameter} is discretized as
\begin{equation}
    \label{eq:diameterUpdate}
    \frac{\Phi_j^{n+1}-\Phi_j^n}{\Delta t} = -2\mu_j V_{SiC},
\end{equation}
where the superscript $n$ represents a value at the current time $t$, $n+1$ represents a value at the next time level $t+\Delta t$, etc. Reaction rate $\mu_j$ is calculated using the $MTS$ concentration at grid location $j$. Lastly, total molar flux $\widehat{N}$ updates as
\begin{equation}
    \widehat{N} = \mu \widehat{s}.
\end{equation}

Before infiltration, it is assumed the pore has a uniform diameter, $\Phi_0 = 2r_{p0}$, along the longitudinal direction. A Dirichlet boundary condition is considered at $z=0$, and thus $C_i(0,t)$ is given and will be provided in Sec.~\ref{sec:CVI-Results}. At $z=L$ a Neumann boundary condition is applied, $\partial C_i(L,t)/\partial z=0$.

\begin{figure}[ht]
    \centering
    \includegraphics[width=0.85\linewidth]{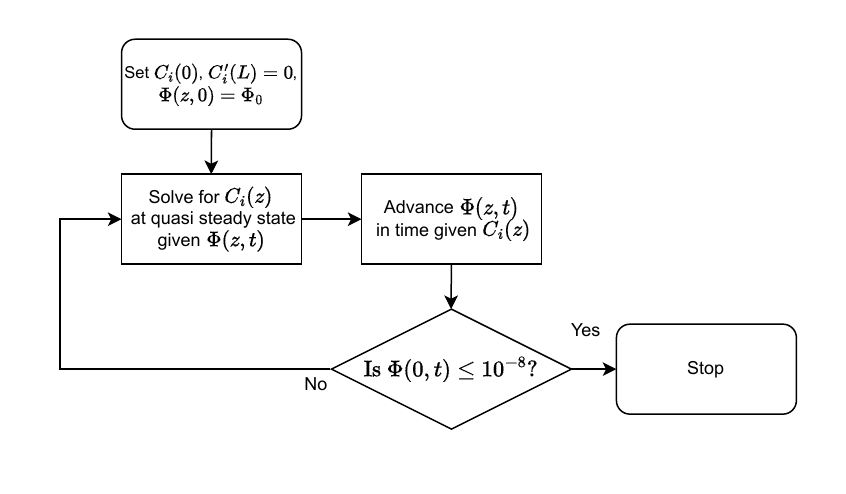}
    \caption{\doublespacing The algorithm coupling gas concentration and pore size.}
    \label{fig-cvimodelingalgorithm}
\end{figure}

The general solution method is outlined in Fig.~\ref{fig-cvimodelingalgorithm}. Given a current pore diameter, $\Phi(z,t)$, the quasi-steady concentration profile, $C_i(z)$, is obtained via Eq.~\eqref{nonlinear-discretized}. Using the updated $MTS$ concentration, $C_{MTS}$, the reaction rate, Eq.~\eqref{reaction-rate}, is used to update the local pore diameter via Eq.~\eqref{eq:diameterUpdate}. If the pore diameter at $z=0$ is small enough the iteration is concluded. Otherwise a step forward in time is taken.

\subsection{Porosity to properties}\label{Sec2.3}

Once the CVI process concludes the shape of the pore radius, $r_p=\Phi/2$, across the length of $z$ is used to create a cylindrical representative volume element, or RVE, of radius $R_V$. This RVE is used to describe the heterogeneous pore microstructure in relationship to the entire $SiC$ specimen~\cite{Chateau2015}. A scalar order parameter for damage due to porosity, $\xi \in [0,1]$, is defined by the ratio of the pore volume compared to the RVE volume at a given location,
\begin{equation}\label{Eq11}
    \xi(z)|_{t=0} =\frac{\pi r_{p}^2(z) L}{\pi R_V^2L} = \frac{r_p^2(z)}{R_V^2}.
\end{equation}
Note that the pore's tortuosity, which increases the path length of the transported species, is treated here as negligible. The tortuosity, $\tau$, of an randomly constructed porous medium can be estimated as a logarithmic function of the preform porosity \cite{Kim2021,Matyka2008}, the path length calculated as $\tau L$ \cite{Ghanbarian2013}, and the RVE volume calculated as a function of the path length. Other complex pore geometries are studied seperately in Guan and Cha \cite{Guan2013-2,Cha2022}, and concave inter-bundle pores are studied in Kulik \cite{Kulik2004}. Interested readers in pore modeling can refer to Vignoles \cite{Vignoles2015}, who reviewed various other modeling strategies of CVI.

What follows is a model of damage evolution demonstrated first by Sun et al~\cite{Sun2022-1} and summarized here. Interested readers in nonlinear phase field modeling can refer to \cite{Chen2015-2, Chen2014, Chen2015-3}. Damage variable $\xi$ is used to define a quadratic degradation function~\cite{Sargado2018}
\begin{equation}\label{Eq12}
    \omega = (1-\xi)^2,
\end{equation}
which smoothly evolves inversely to $\xi$ such that $\omega(1) = 0$ and $\omega(0) = 1$. This degradation function is then used to model the effect of porosity on the Young's modulus, $E$, and Poisson's ratio, $\mu$~\cite{Perkins2020}.

The evolution of $\xi$ over time is governed by the Allen-Cahn equation~\cite{Allen1972}
\begin{equation}\label{Eq13}
\frac{\partial \xi}{\partial t} = -M\biggl(\frac{\partial \Psi(\xi)}{\partial \xi} - \frac{\partial }{\partial x_i}\biggl(  \gamma \frac{\partial \xi}{\partial x_i}    \biggl)\biggl),
\end{equation}
where $M$ is a coefficient of mobility, $\Psi$ constitutes local free energy, and $\gamma$ is the coefficient of the surface energy contribution from $\xi$. The local free energy, $\Psi$, can be decomposed into the contributions from elasticity and from damage such that
\begin{equation}
    \Psi = \Psi_{\text{elas}} + \Psi_{\text{frac}}.
\end{equation}
The material upon fracture can withstand compressive stress but not tensile stress. Therefore, it is assumed that the damage only affects the tensile strain and stress. Through strain spectral decomposition, the elastic strain energy $\psi$ is separated into tensile and compressive components $\psi^+$ and $\psi^-$. That is,
\begin{equation}
    \psi= \psi^+ + \psi^-.
\end{equation}
Crack irreversibility is a necessary condition as the crack does not self heal. A history variable,
\begin{equation}
    \mathcal{H} = \max_t{(\psi^+, \Psi_\text{b})},
\end{equation}
is then introduced to identify the maximum tensile strain energy across time, where the free energy lower bound~\cite{Chukwudozie2019,Miehe2015,Mikelic2019,Wu2017}
\begin{equation}
    \Psi_\text{b} = -\frac{\Psi_{\text{frac}}}{\omega(\xi)}\Biggl|_{\xi=0}
\end{equation}
ensures $\mathcal{H}\geq 0$. Then, the elastic free energy is~\cite{Sun2022-1}
\begin{equation}
    \Psi_{\text{elas}} = \omega(\xi)\mathcal{H}+\psi^{-}.
\end{equation}
Modeling the volumetric fracture energy term as a quadratic expression is an approach documented by Miehe et al~\cite{Miehe2010, Miehe2015}. The corresponding fracture free energy terms are
\begin{equation}\label{Eq19}
    \Psi_{\text{frac}} = \frac{g_c\xi^2}{2l_0}, \qquad \gamma = \frac{g_cl_0}{2},
\end{equation}
where $g_c$ is the critical energy release rate defined by Griffith's theory of fracture~\cite{Lajtai1971} and $l_0$ is the length scale of the physical crack. As a means for calculating $g_c$ and $l_0$, Sec. \ref{Sec:gcl0} is provided.
\subsubsection{Energy release rate and crack length scale}\label{Sec:gcl0}
The justification behind the relationship between $l_0$ and $\Psi_{\text{frac}}$ in Eq.~\eqref{Eq19} is as follows. Surface cracks can be approximated in 2D with the diffuse function
\begin{equation}
\xi(x) = e^{-|x|/l_0},
\end{equation}
which has boundary conditions
\begin{equation}
\xi(0) = 1, \ \ \ \ \  \xi(\pm\infty) = 0
\end{equation}
and is a solution to the second order ordinary differential equation
\begin{equation}
-\xi''(x) + \frac{1}{l_0^2}\xi (x) = 0.
\end{equation}
Using the principle of virtual work, its Galerkin weak form is
\begin{equation}
A(\xi) = \int_{-\infty}^\infty\biggl[  \frac{\xi^2}{2 l_0} + \frac{l_0 |\nabla \xi |^2}{2} \biggl] dV,
\end{equation}
which has been previously seen~\cite{Miehe2010, Zhang2017} and supports the idea of Moln\'ar et al that the fracture free energy is quadratic~\cite{Molnar2020}. With $A(\xi)$ as the integral of the ODE, it also represents the surface area under the curve that characterizes the crack shape. This means that the energy release rate $g_c$ can be scaled by surface area $A$ to compute
\begin{equation}\label{Eq24}
\Gamma = g_c A = g_c \int_{-\infty}^{\infty}\Biggl[  \frac{\xi^2}{2 l_0} + \frac{l_0 |\nabla \xi |^2}{2} \Biggl] dV,
\end{equation}
which represents the energy release across the area of the body necessary to generate a new crack. Despite $g_c$ and $l_0$ still being classified so far as unknowns in Eq.~\eqref{Eq19} and Eq.~\eqref{Eq24}, they can be solved analytically provided governing equations and boundary conditions associated with simple tension~\cite{Miehe2015, Zhang2017, Kuhn2015, Tanne2018}. Their solutions are
\begin{equation}\label{Eq25}
l_0 = \frac{27}{256}\frac{K_{Ic}^2}{\sigma_c^2}, \ \ \ \ \ g_c = \frac{K_{Ic}^2}{E},
\end{equation}
where $K_{Ic}$ is the critical stress concentration factor under mode I fracture, $E$ is the elastic modulus, and $\sigma_c$ is the applied critical stress associated with $K_{Ic}$, all of which are well-documented for many materials. With these values being known, they can be applied to the fracture energy terms in Eq.~\eqref{Eq19}, which are crucial inputs to the equation characterizing the evolution of damage over time, which is Eq.~\eqref{Eq13}.

\subsection{Properties to performance}\label{properties-performancesection}
Using the finite element solver MOOSE~\cite{Permann2020}, the formulation summarized in Sec. \ref{Sec2.3} can be used to model any mode I fracture problem. One of particular interest is the four point bending test, out of which values for flexural strength can be obtained. The simulation setup is for a standard four point bending test specified in ASTM C1161-18~\cite{C282013}. A slender rectangular specimen is generated, with a length that is approximately 20 times the height. Plane stress conditions are assumed, and elements are uniformly distributed. The element size must be at least ten times smaller than the crack length scale, $l_0$, in order to adequately resolve the crack length (various FE mesh discretizations are tried in Sec. \ref{sec-femesh}). Various $l_0$ are shown in Table \ref{table-l0} and are based on Eq.~\eqref{Eq25}.

\begin{table}[ht]
\centering
\resizebox{0.65\columnwidth}{!}{%
\begin{tabular}{llllll}
$T_{\text{FPB}}$ $\text{(K)}$ & $K_{IC}$ $\text{(m}^{1/2}\text{MPa)}$ & $E$ $\text{(MPa)}$ & $\sigma_c \ \text{(MPa)}$  & $g_c$ $\text{(J}/\text{m}^2\text{)}$ & $l_0 \ (\mu \text{m)}$ \\ \hline
293.15 & 3.88 & 0.414 & 250 & 36.32 & 25.4 \\
473.15 & 3.74 & 0.410 & 250 & 34.08 & 23.6 \\
673.15 & 3.54 & 0.406 & 250 & 30.88 & 21.1 \\
873.15 & 4.04 & 0.401 & 250 & 40.68 & 27.5 \\
1073.15 & 4.27 & 0.397 & 250 & 45.97 & 30.8 \\
1273.15 & 4.19 & 0.392 & 250 & 44.79 & 29.6 \\
1473.15 & 3.95 & 0.387 & 250 & 40.27 & 26.3 \\
1673.15 & 4.73 & 0.383 & 250 & 58.45 & 37.8
\end{tabular}
}
\caption{\doublespacing Temperature-dependent material properties of $\alpha$-$SiC$~\cite{Sun2022-1}.}
\label{table-l0}
\end{table}

 Two support points are set at the bottom of the specimen and two loading points are set at the top surface of the specimen, Fig.~\ref{fig-cvifs}. A Dirichlet displacement boundary condition is applied downward onto the top of the body. In a bending test, it is assumed that the specimen is essentially one-dimensional, since the fracture is expected to propagate in the $z$-direction only. Because of the phase field mapping done using Eq.~\eqref{Eq11}, the damage curve must always be contained between 0 and  $r^2_{p0}/R_V^2$.

Note that because porosity is direction-dependent, the material strength from the top surface is different from that of the bottom. Flexural strength, $\sigma_f$, is defined as the maximum stress that the specimen sustains, which takes place at the bottom surface, and this is a variable that will be focused on in the following sections.

Work in MOOSE is undergirded by a math library provided by PETSc~\cite{PETSc2022-1,PETSc2022-2,PETSc2022-3,PETSc2022-4}, which offers a variety of solver and preconditioner options. The combination of an iterative PJFNK solver and a BoomerAMG preconditioner are optimal for solving the Allen-Cahn equation for phase field fracture, and are therefore utilized for this problem. Scalability characteristics of the framework, as well as additional commentary on the stability and reliability of different executioner options, are found in ~\cite{Sun2022-1}.

\section{Results/discussion}\label{Sec3}
\subsection{Convergence testing}\label{sec-convergence}
The purpose of this section is to ensure the reliability of the numerical model using convergence tests.
As there are no closed-form solutions that can be compared against, in all cases results will be compared to a fine mesh/small time step to ensure that the methods are converging to a single result. The time step of the CVI process is denoted by $\Delta t_{\text{CVI}}$. The grid spacing of the CVI process is $\Delta z_{\text{CVI}} = L/(N-1)$, where $L$ is the pore length and $N$ is the number of grid points. The time step of the bending test is $\Delta t_{\text{FPB}}$, and the FE mesh discretization of the bending test specimen is $\Delta z_{\text{FPB}} = L/N_e$, where $N_e$ is the number of elements in the FE mesh. As seen in Fig. \ref{fig-cvifs}, the height of the mesh is set equal to the pore length. The convergence behavior of $L=\{5, \ 0.075\}$ mm are compared in Secs. \ref{sec-timeconvergence}-\ref{sec-coupledconvergence}. The first pore length is further used in Sec. \ref{sec-validation} as a means of model validation with respect to the literature. The second pore length is further used in the four point bending tests in Sec. \ref{sec-flexureresponse} because it is approximately twice the crack length scale.

\subsubsection{Temporal convergence}\label{sec-timeconvergence}
\begin{figure}
\centering
\begin{subfigure}{.5\textwidth}
  \centering
  \includegraphics[width=\linewidth]{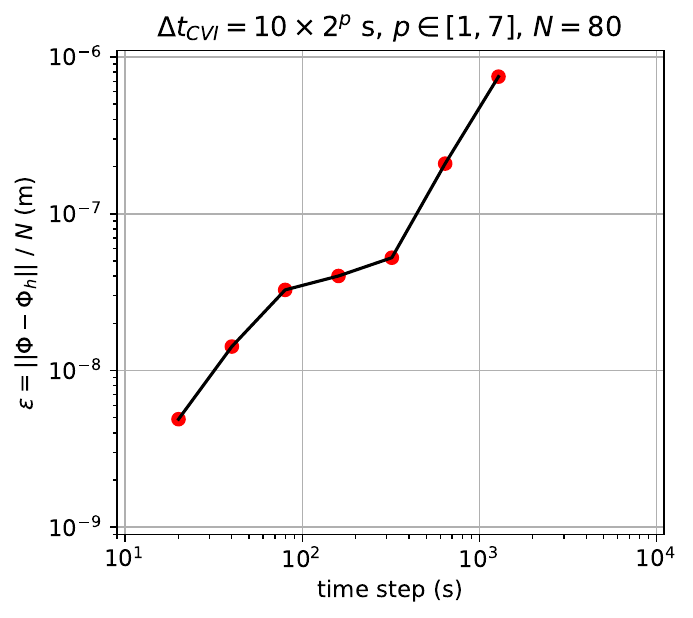}
  \caption{\doublespacing $L=5$ mm.}
  \label{fig-temporalLbig}
\end{subfigure}%
\begin{subfigure}{.5\textwidth}
  \centering
  \includegraphics[width=\linewidth]{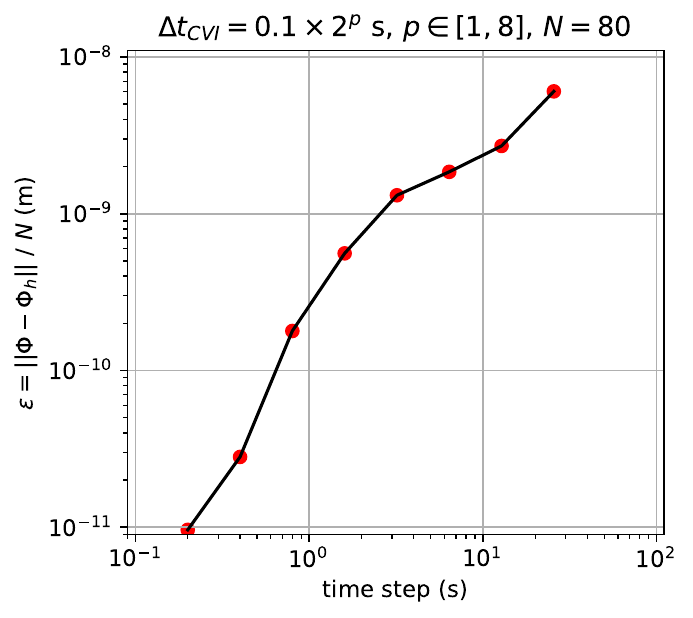}
  \caption{\doublespacing $L=0.075$ mm.}
  \label{fig-temporalLsmall}
\end{subfigure}
\caption{\doublespacing Average $L_{2}$ norm of the difference in pore profile between tests with time steps $\Delta t_{\text{CVI}} = \Delta t_{base} \times 2^p$ and a test with $\Delta t_{\text{CVI}}=\Delta t_{base}$.}
\label{fig:dtloglog}
\end{figure}

As is well known, there is a trade-off between a large (faster solve time) and a small (higher accuracy) time step.
A large time step improves wall clock time as it solves for the system of gas concentration and pore size PDEs faster at the expense of accuracy. On the other hand, a smaller time step ensures greater accuracy, because finer changes in the gas concentration and pore profiles are accounted for, but take longer to complete. Here time steps of $\Delta t_{\text{CVI}} = \Delta t_{base} \times 2^p$ seconds are compared to the result obtained using $\Delta t_{base}$ seconds, with the average $L_2$-norm of the profile difference along the entire depth shown in Fig.~\ref{fig:dtloglog} (in which constants are $T_{\text{CVI}}=1273$ K, $P=20$ kPa, $\alpha=1$, $L/\Phi_0=50$). The factor $p=1,2,3,\dotsc$  doubles the time step in subsequent cases. For $L=5$ mm the baseline is $\Delta t_{base}=10$ seconds and for $L=0.075$ mm the baseline is $\Delta t_{base}=0.1$ seconds. The convergence for both cases are slightly higher than first-order: for $L=5$ mm it is $\sim$ 1.07, and for $L=0.075$ mm it is $\sim$ 1.30. This is expected as the time discretization is first-order accurate. Table \ref{table-solvetime-time} shows the solve time versus the time step.

\FloatBarrier

\begin{table}[ht]
\centering
\begin{tabular}{llllll}
$L$ (mm) & $\Delta t_{\text{CVI}}$ (s) & CVI Solve time (s)
\\ \hline
$5$  & $10=\Delta t_{base}$           & $29.22$
\\
& $20$           & $14.55$
\\
& $40$           & $7.41$
\\
& $80$           & $3.79$
\\
& $160$           & $1.90$
\\
& $320$           & $0.98$
\\
& $640$           & $0.59$
\\
& $1280$           & $0.38$
\\
           &
\end{tabular}
\begin{tabular}{llllll}
$L$ (mm) & $\Delta t_{\text{CVI}}$ (s) & CVI Solve time (s)
\\ \hline
$0.075$ & $0.1=\Delta t_{base}$           & $43.41$
\\
& $0.2$           & $21.47$
\\
& $0.4$           & $10.80$
\\
& $0.8$           & $5.47$
\\
& $1.6$           & $2.84$
\\
& $3.2$           & $1.54$
\\
& $6.4$           & $0.76$
\\
& $12.8$           & $0.40$
\\
& $25.6$           & $0.19$
\end{tabular}

\caption{\doublespacing CVI solve time vs. time step for $L=\{5, \ 0.075\}$ mm, $L/\Phi_0=50$.}
\label{table-solvetime-time}
\end{table}

\FloatBarrier

\subsubsection{Grid convergence}\label{sec-gridconvergence}
\begin{figure}
\centering
\begin{subfigure}{.5\textwidth}
  \centering
  \includegraphics[width=\linewidth]{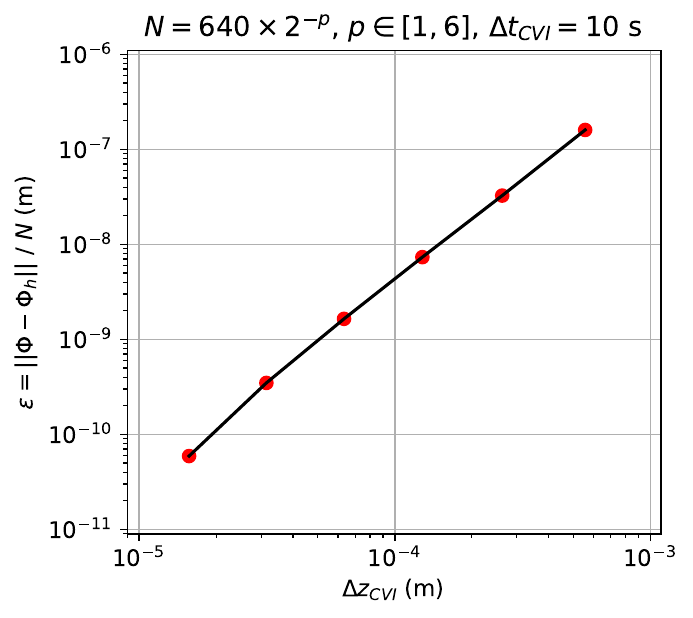}
  \caption{\doublespacing $L=5$ mm.}
  \label{fig-spatialLbig}
\end{subfigure}%
\begin{subfigure}{.5\textwidth}
  \centering
  \includegraphics[width=\linewidth]{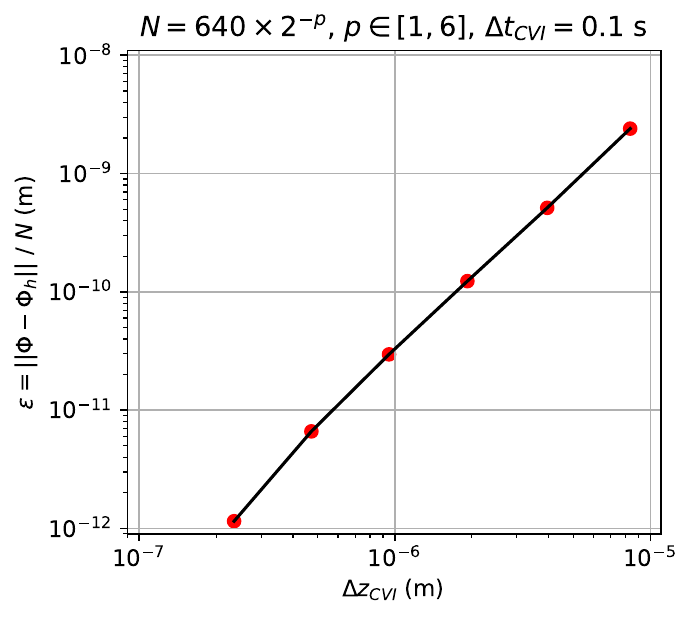}
  \caption{\doublespacing $L=0.075$ mm.}
  \label{fig-spatialLsmall}
\end{subfigure}
\caption{\doublespacing Average $L_{2}$ norm of the difference in pore profile between tests with grid spacings $\Delta z_{\text{CVI}} = L/(N_{base}\times 2^{-p} -1)$ and a test with $\Delta z_{\text{CVI}}= L/(N_{base}-1)$, where $N_{base}=640$.}
\label{fig:Nloglog}
\end{figure}

In addition to the time step, the grid spacing needs to be investigated to ensure that the relevant physics are being modelled correctly with the least amount of computational cost as
the ideal grid yields an accurate solution while possessing the fewest possible grid points. In this section the pore profile is obtained using a time step of $\Delta t=\Delta t_{base}$ seconds for grid spacings of $\Delta z_{\text{CVI}}=L/(N-1)$, where $N = N_{base} \times 2^{-p}$. The factor $p=1,2,\dotsc, 6$ halves the grid spacing in subsequent tests. These are compared to the result obtained using $N=N_{base}=640$. As before the average $L_2$-norm is calculated and reported in Fig.~\ref{fig:Nloglog} (in which constants are $T_{\text{CVI}}=1273$ K, $P=20$ kPa, $\alpha=1$, $L/\Phi_0=50$). The convergence for both cases are approximately second-order: for $L=5$ mm it is $\sim$ 2.19, and for $L=0.075$ mm it is $\sim$ 2.11. The spatial discretization is second-order accurate, so this is expected. Table \ref{table-solvetime-N} compares the solve time to the number of grid points.

\FloatBarrier

\begin{table}[ht]
\centering
\begin{tabular}{llllll}
$L$ (mm) & $N$ & CVI Solve time (s)
\\ \hline
$5$  & $10$           & $0.41$
\\
& $20$           & $1.86$
\\
& $40$           & $7.43$
\\
& $80$           & $29.68$
\\
& $160$           & $124.63$
\\
& $320$           & $569.11$
\\
& $640 = N_{base}$           & $2821.00$
\end{tabular}
\begin{tabular}{llllll}
$L$ (mm) & $N$ & CVI Solve time (s)
\\ \hline
$0.075$  & $10$           & $0.63$
\\
& $20$           & $2.71$
\\
& $40$           & $11.00$
\\
& $80$           & $43.99$
\\
& $160$           & $168.93$
\\
& $320$           & $857.71$
\\
& $640 = N_{base}$           & $4205.11$
\end{tabular}

\caption{\doublespacing CVI solve time vs. grid points for $L=\{5, \ 0.075\}$ mm, $L/\Phi_0=50$.}
\label{table-solvetime-N}
\end{table}

\FloatBarrier

\subsubsection{Coupled convergence}\label{sec-coupledconvergence}
\begin{figure}
\centering
\begin{subfigure}{.5\textwidth}
  \centering
  \includegraphics[width=\linewidth]{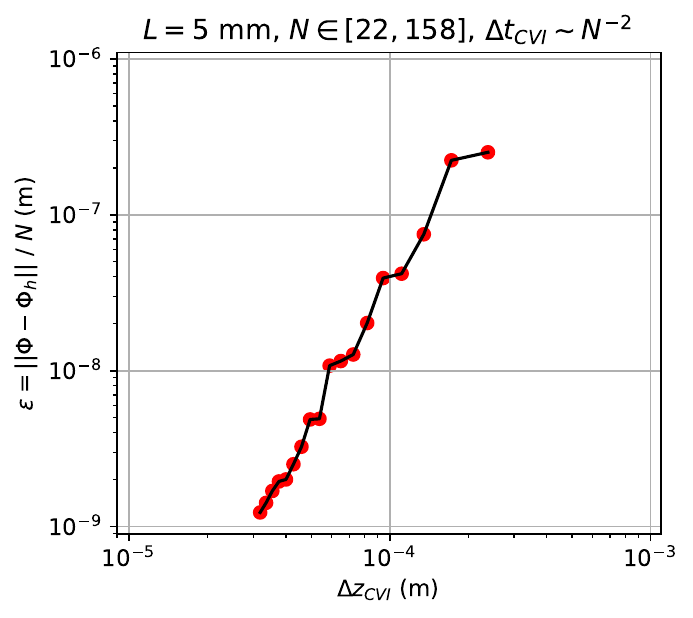}
  \caption{\doublespacing $L=5$ mm.}
  \label{fig-NsquaredLbig}
\end{subfigure}%
\begin{subfigure}{.5\textwidth}
  \centering
  \includegraphics[width=\linewidth]{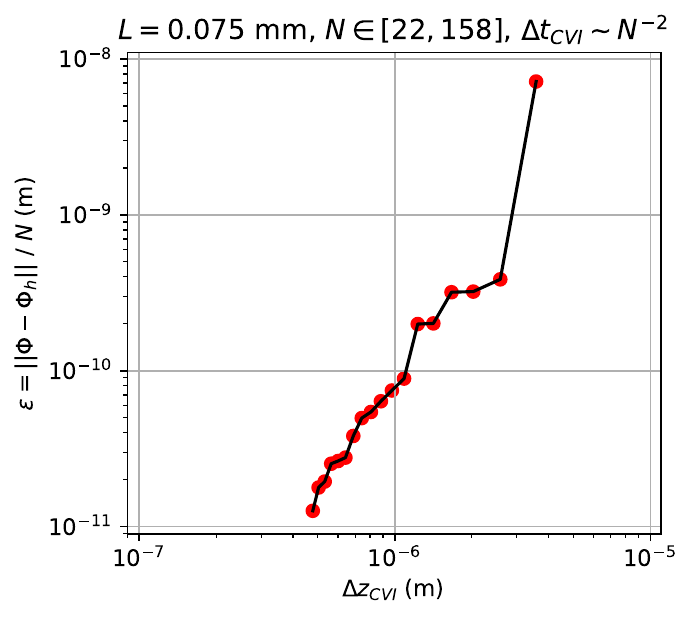}
  \caption{\doublespacing $L=0.075$ mm.}
  \label{fig-NsquaredLsmall}
\end{subfigure}
\caption{\doublespacing Average $L_2$ norm of the difference in pore profile between grids with $N\in[22, 158]$ and a grid with $N=160$. In all cases, $\Delta t_{\text{CVI}} \sim N^{-2}$, $\Delta z_{\text{CVI}} = L/(N-1)$. }
\label{fig:Nsquaredloglog}
\end{figure}

To explore the coupled nature of the CVI model the result for $\Delta t_{\text{CVI}} = \Delta t_{base}$ and $\Delta z_{\text{CVI}} = L/(160-1)$ is compared to the results obtained for $\Delta z_{\text{CVI}} = L/(N-1)$, $N\in[20, 158]$ with $\Delta t_{\text{CVI}} \sim N^{-2}$, with the results shown in Fig.~\ref{fig:Nsquaredloglog} (in which constants are $T_{\text{CVI}}=1273$ K, $P=20$ kPa, $\alpha=1$, $L/\Phi_0=50$). The order of convergence for $L=5$ mm is calculated to be $\sim 2.86$, and for $L=0.075$ mm is calculated to be $\sim 2.52$. Table \ref{table-solvetime-Nquared} shows the solve time versus the numerical parameters.

\begin{table}[ht]
\centering
\begin{tabular}{llllll}
$L$ (mm) & $N$ & $\Delta t_{\text{CVI}}$ (s) & CVI solve time (s)
\\ \hline
$5$ & $22$  & $528.93$ & $0.04$
\\
& $30$  & $284.44$ & $0.17$
\\
& $38$  & $177.29$ & $0.42$
\\
& $46$  & $120.98$ & $0.84$
\\
& $54$  & $87.79$ & $1.61$
\\
& $62$  & $66.60$ & $2.63$
\\
& $70$  & $52.24$ & $4.46$
\\
& $78$  & $42.08$ & $6.96$
\\
& $86$  & $34.61$ & $10.49$
\\
& $94$  & $28.97$ & $15.02$
\\
& $102$  & $24.61$ & $20.56$
\\
& $110$  & $21.16$ & $29.55$
\\
& $118$  & $18.39$ & $37.49$
\\
& $126$  & $16.12$ & $49.11$
\\
& $134$  & $14.26$ & $62.96$
\\
& $142$  & $12.70$ & $80.07$
\\
& $150$  & $11.38$ & $100.48$
\\
& $158$  & $10.25$ & $124.88$
\\
& $160$  & $10.00$  & $131.80$
\end{tabular}
\begin{tabular}{llllll}
$L$ (mm) & $N$ & $\Delta t_{\text{CVI}}$ (s) & CVI solve time (s)
\\ \hline
$0.075$ & $22$  & $5.29$ & $0.06$
\\
& $30$  & $2.84$ & $0.19$
\\
& $38$  & $1.77$ & $0.53$
\\
& $46$  & $1.21$ & $1.12$
\\
& $54$  & $0.88$ & $2.23$
\\
& $62$  & $0.67$ & $3.91$
\\
& $70$  & $0.52$ & $6.28$
\\
& $78$  & $0.42$ & $9.95$
\\
& $86$  & $0.35$ & $14.55$
\\
& $94$  & $0.29$ & $21.15$
\\
& $102$  & $0.25$ & $29.50$
\\
& $110$  & $0.21$ & $39.96$
\\
& $118$  & $0.18$ & $52.98$
\\
& $126$  & $0.16$ & $69.87$
\\
& $134$  & $0.14$ & $90.62$
\\
& $142$  & $0.13$ & $115.32$
\\
& $150$  & $0.11$ & $144.46$
\\
& $158$  & $0.10$ & $179.10$
\\
& $160$  & $0.10$  & $188.99$
\end{tabular}

\caption{\doublespacing CVI solve time vs. coupled spacing for $L=\{5, \ 0.075\}$ mm, $L/\Phi_0=50$.}
\label{table-solvetime-Nquared}
\end{table}

\FloatBarrier

\subsubsection{FE mesh convergence}\label{sec-femesh}
This section determines an appropriate FE mesh spacing, $\Delta z_{\text{FPB}} = L/N_e$, and time step, $\Delta t_{\text{FPB}}$, for the four point bending tests in Sec. \ref{sec-flexureresponse}. For all cases, $L$ is the height of the specimen and $N_e$ is the number of vertical finite elements.

Using Cubit, the specimen described in Sec. \ref{properties-performancesection} is generated. The temperature of the four point bending environment is $T_{\text{FPB}}=1073.15$ K, which corresponds to a crack length scale of $l_0=30.8 \ \mu$m (Table \ref{table-l0}). To ensure that the crack is adequately resolved in the mesh, the spacing must be at least ten times smaller than the crack length scale, $30.8 \ \mu \text{m}/10 = 3.08 \ \mu$m. For all cases, the height of the specimen is $L=0.075$ mm $=75 \ \mu$m, because this is roughly twice the length of the expected crack. Therefore, the minimum number of elements along the height of the FE mesh is set to $N_e=28$, so that $\Delta z_{\text{FPB}}=75 \ \mu$m/28 $\approx 2.67 \ \mu$m $< 3.08 \ \mu$m. In subsequent tests the mesh is further refined and the change in flexural strength is observed.

A maximum element number of $N_e=448$ is chosen as the baseline. Cases with $N_e = 448$ and time steps $\Delta t_{\text{FPB}}= 5, \ 10, \ 20$ ms are compared. The percent difference in $\sigma_f$ between the 5 ms case and the 20 ms case is 0.31\%. The difference between the 5 ms case and the 10 ms case is a negligible 0.10\%. Therefore, a time step of 10 ms is adopted for testing.

Table \ref{table-hcases} shows the results corresponding to various mesh resolutions (in which constants are  $T_{\text{CVI}}=1223$ K, $P=20$ kPa, $L=0.075$ mm, $L/\Phi_0$ = 50, $\alpha=1$, 30\% initial porosity,
$T_{\text{FPB}}=1073$ K, $\Delta t_{\text{FPB}}=10$ ms). The percent difference between the flexural strength of the coarse grid ($N_e=28$) case and the fine grid ($N_e=448$) case is $0.11$\%. The $L_{\infty}$ norm between the two cases is negligible. Neither the flexural strength nor the time to fracture are significantly affected by the mesh element spacing. Therefore it is acceptable to use $\Delta z_{\text{FPB}}=L/N_e, \ N_e=28$ as the grid spacing in the $z$ direction in the finite element mesh. The slope of the convergence is $1.21$, indicating approximately a first order accurate numerical method.

\begin{table}[ht]
\centering
\begin{tabular}{llllll}
$N_e$ & $\sigma_f$ (MPa) & $t_{\sigma_f}$ (s)

\\ \hline
$28$           & 374.332 & 4.99
\\
$112$          & 374.653 & 4.99
\\
$224$ & 374.729 & 4.98
\\
$448$ & 374.761 & 4.98
\end{tabular}

\caption{\doublespacing Flexural strength ($\sigma_f$) and time to fracture ($t_{\sigma_f}$) for different mesh resolutions.}
\label{table-hcases}
\end{table}

\subsection{CVI-reduced pore distributions}\label{sec-validation}
\label{sec:CVI-Results}

This section validates the model with respect to the surrounding literature, in which a pore length of $L=5$ mm is commonly used. Based on the convergence testing in Sec. \ref{sec-convergence}, all tests in this section utilize a time step of $\Delta t_{\text{CVI}} = 10$ s and a grid spacing of $\Delta z_{\text{CVI}} = L/(N-1)$, where $N=80$ is the number of grid points. This combination is chosen because of its ability to rapidly calculate solutions while not sacrificing more than an $\varepsilon$ of $10^{-8}$ m (Fig. \ref{fig-spatialLbig}).

\begin{figure}
\centering
\begin{subfigure}{.5\textwidth}
  \centering
  \includegraphics[width=\linewidth]{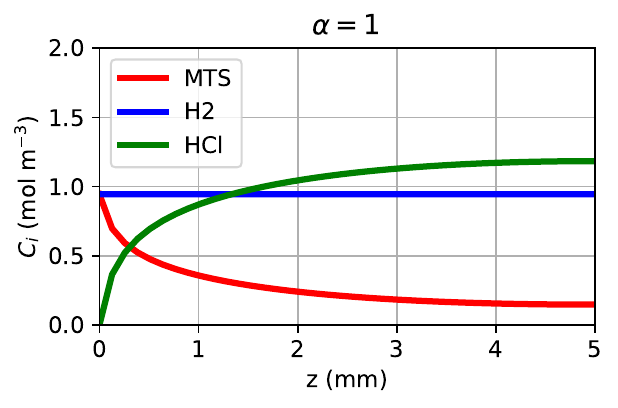}
  \caption{\doublespacing $C_{MTS,0}=C_{H2,0}$.}
  \label{fig-alpha1}
\end{subfigure}%
\begin{subfigure}{.5\textwidth}
  \centering
  \includegraphics[width=\linewidth]{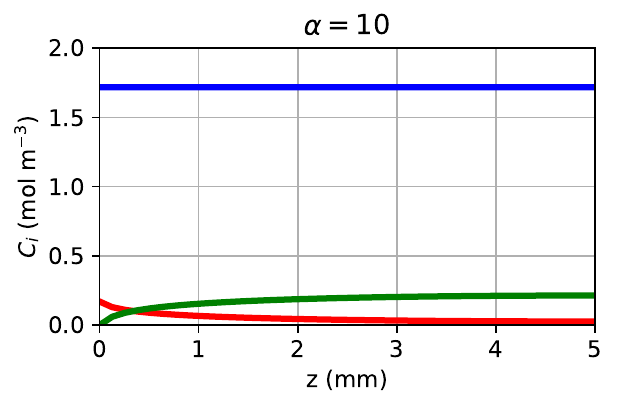}
  \caption{\doublespacing $10C_{MTS,0}=C_{H2,0}$.}
  \label{fig-alpha10}
\end{subfigure}
\caption{\doublespacing Final gas concentration profiles with gas ratio $\alpha$.}

\end{figure}

The quantities of interest in the examination of a CVI process is its infiltration depth, $d$, and time of infiltration, $t_f$. Infiltration depth $d = (\Phi_0-\Phi)/2 = r_{p0}-r_p$ is defined as the amount of $SiC$ matrix material deposited or equivalently the degree of pore shrinkage~\cite{Fedou1993,Fedou1993-2}. Time of infiltration, $t_f$, is defined as the amount of time required for the pore diameter at the surface to close. These two quantities are dependent on the physical phenomena of pore filling and gas transport. The CVI process is modeled as a function of the process parameters of temperature $T_{\text{CVI}}$, pressure $P$, and the gas ratio $\alpha = C_{H2,0}/C_{MTS,0}$. At the reaction interface, $\alpha$ is enforced by the boundary conditions
\begin{equation}
    C_{MTS,0} = \frac{1}{1+\alpha}\frac{P}{RT_{\text{CVI}}}, \ \ \ \ C_{H2,0} = \frac{\alpha}{1+\alpha}\frac{P}{RT_{\text{CVI}}}, \ \ \ \ C_{HCl,0} = 0,
\end{equation}
where $R$ is the ideal gas constant. Isobaric and isothermal conditions are assumed for each test. The enforcement of these boundary conditions is demonstrated in Figs. \ref{fig-alpha1}, \ref{fig-alpha10} (in which constants are $T_{\text{CVI}}=1273 \text{ K}$, $P=20 \text{ kPa}$, $\Phi_0=100 \ \mu \text{m}$, $L=5\text{ mm}$).
At $z=0$, $C_i$ does not change from its initial magnitude. The gas ratio $\alpha=1$ leads to $ C_{H2,0}=C_{MTS,0}$. Therefore, the concentrations at the infiltration interface are equal, as shown in Fig. \ref{fig-alpha1}. However a gas ratio of $\alpha=10$ implies $ C_{H2,0}=10C_{MTS,0}$. As a result, there is no intersection in Fig. \ref{fig-alpha10}. $H_2$ remains constant as a carrier gas. Therefore, it has equal weight on the side of the reactants and that of the products. This motivates the stoichiometric coefficient $s_1 = s_{H2}=0.$ The other coefficients are $s_{MTS}=-1$ and $s_{HCl}=3$ because of the sign convention of positive products and negative reactants. This sign convention motivates the increase in $C_{HCl}$ over time and across the depth of the substrate, while $MTS$ decomposes and diminishes.

\begin{figure}[ht]
    \centering
    \includegraphics[width=0.85\linewidth]{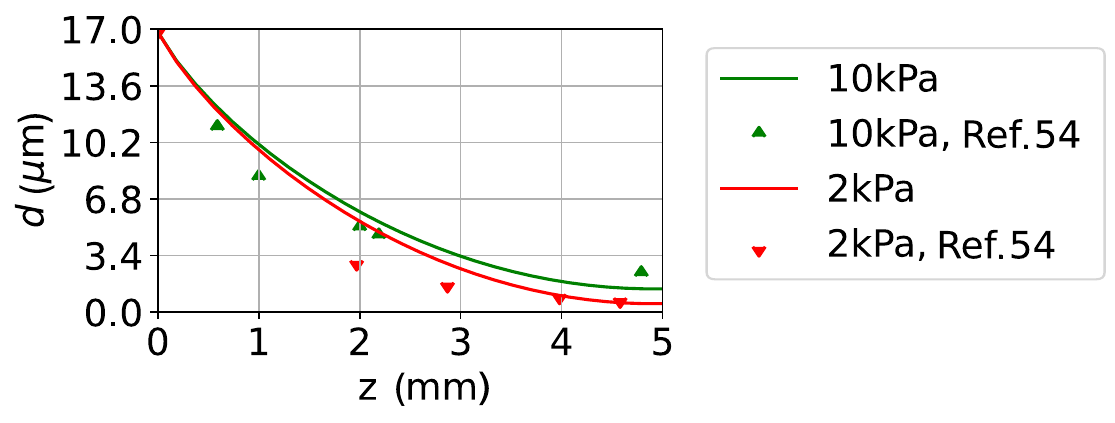}
    \caption{\doublespacing Final infiltration depth profiles with CVI pressures 2 kPa and 10 kPa compared to Ref.~\cite{Fedou1993}.}
    \label{fig-d2-10kpa}
\end{figure}

For purposes of validation, it is necessary to compare the current model with results gathered in the literature. \begin{figure}[ht]
    \centering
    \includegraphics[width=0.7\linewidth]{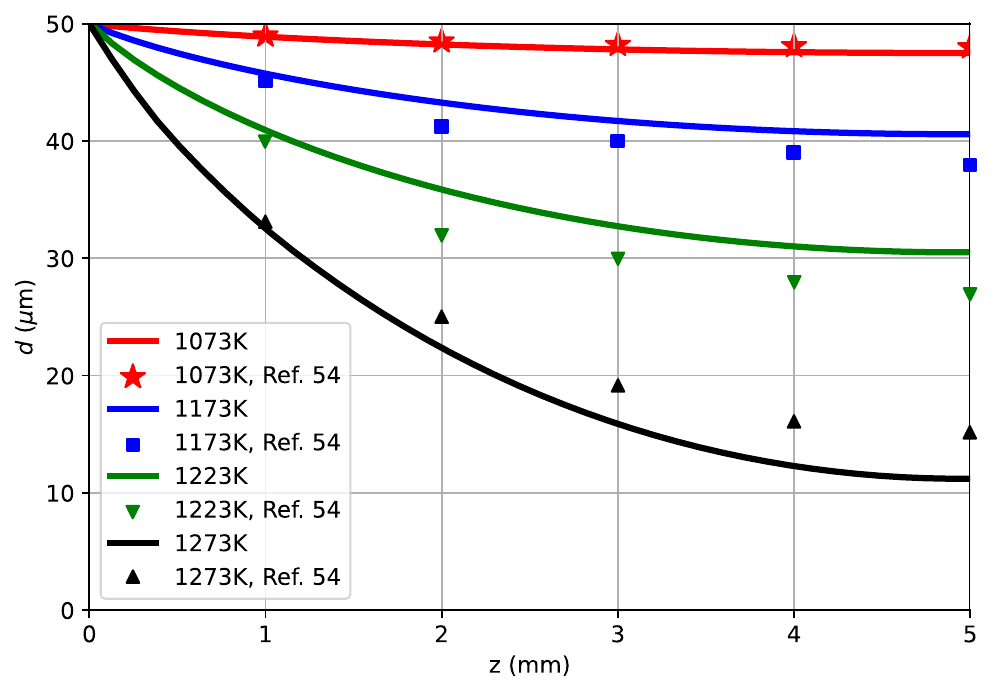}
    \caption{\doublespacing Final infiltration depth profiles with CVI temperatures 1073 K, 1173 K, 1223 K, and 1273 K compared to Ref.~\cite{Fedou1993}.}
    \label{fig-d1073-1273}
\end{figure}
The results of this study fall within 10\% of the data reported in Ref.~\cite{Fedou1993}. In Fig. \ref{fig-d1073-1273} (in which constants are $P=20\text{ kPa}$, $\alpha=$ 10, $\Phi_0=100 \ \mu \text{m}$, $L=5\text{ mm}$), the infiltration depth is sensitive to changes in temperature. This is due to the high activation energy of the CVI process being modeled (Sec. \ref{sec-processingporosity}). For chemical reactions with high activation energies, variations in temperature significantly impact the progress of the chemical reaction. This physically observed behavior motivates the modeling strategy, in which reaction rate exponentially increases with temperature. In Fig. \ref{fig-d2-10kpa} (in which constants are $T_{\text{CVI}}=1223\text{ K}$, $\alpha=5$, $\Phi_0=34 \ \mu \text{m}$, $L=5\text{ mm}$), the final pore profile is relatively pressure-independent due to the length scale of the porous media being experimented on, where Knudsen diffusion is dominant. The Knudsen diffusion coefficient depends only on pore geometry and temperature, not pressure. On the other hand, Fick diffusion depends on pressure but predominates in larger pores.

\FloatBarrier

\subsection{Time of infiltration}
\subsubsection{Temperature and pressure dependence}
An increase of temperature implies an increase in kinetic energy, resulting in more frequent collisions. For this reason, the deposition rate increases with temperature in the Arrhenius form, Eq.~\eqref{reaction-rate}. This highly active reaction leads to a decrease in the amount of time required for the pore to close at the interface. In addition, an increased pressure leads to a shortening of the distance beween gas molecules, causing the overall diffusion coefficient to increase, Eqs.\eqref{fickknudsen}-\eqref{gilliland}. This is demonstrated across a wide range of CVI temperatures and pressures as seen in Table \ref{fig-cvitime-tp} (in which constants are $L=5\text{ mm}$, $L/\Phi_0$ = 50). Comparing this information to Fig. \ref{fig-d1073-1273}, it is clear that infiltration time and penetration depth are directly related quantities. This is expected because a longer infiltration time allows the precursor to react with the region of the substrate furthest away from the interface for a longer period of time. Consequently, if the infiltration time is larger, the final pore profile is uniform along the depth. If infiltration time is smaller, the pore profile varies more widely along the depth.

\begin{table}[ht]
    \centering

    \begin{tabular}{ccccc}

         & & \multicolumn{3}{c}{$T_{\text{CVI}}$ (K)} \\
         & \multicolumn{1}{c|}{} & 1273 & 1173 & 1073\\
        \cline{2-5} \cline{3-5} \cline{4-5} \cline{5-5}
        \multirow{5}{*}{\begin{turn}{90} $P$ (kPa)\end{turn}} & \multicolumn{1}{r|}{50} & 0.17 & 1.69 & 26.25 \\
        & \multicolumn{1}{r|}{35} & 0.24 & 2.43 & 37.50      \\
        & \multicolumn{1}{r|}{20} & 0.42 & 4.24 & 65.56 \\
        & \multicolumn{1}{r|}{10} & 0.85 & 8.47 & 131.11      \\
        & \multicolumn{1}{r|}{5}  & 1.69 & 16.95 & 262.09 \\
        \cline{2-5}
        \multicolumn{5}{c}{$\alpha=1$}
    \end{tabular}
    \qquad
    \begin{tabular}{ccccc}

         &   & \multicolumn{3}{c}{$T_{\text{CVI}}$ (K)} \\
         & \multicolumn{1}{c|}{} & 1273 & 1173 & 1073\\
        \cline{2-5} \cline{3-5} \cline{4-5} \cline{5-5}
        \multirow{5}{*}{\begin{turn}{90} $P$ (kPa)\end{turn}} & \multicolumn{1}{r|}{50} & 0.93 & 9.22 & 144.11\\
        & \multicolumn{1}{r|}{35} &  1.33      & 13.67 &  205.86    \\
        & \multicolumn{1}{r|}{20} & 2.33 & 23.28 & 360.56 \\
        & \multicolumn{1}{r|}{10} &  4.66      & 46.11 &  720.52    \\
        & \multicolumn{1}{r|}{5}  & 9.31       & 92.22 &  1441.03 \\
        \cline{2-5}
        \multicolumn{5}{c}{$\alpha=10$}
    \end{tabular}
    \caption{\doublespacing Time of infiltration, $t_f$ (in hours), based on CVI temperature, pressure, and gas ratio.}
    \label{fig-cvitime-tp}
\end{table}

\subsection{Flexure response}\label{sec-flexureresponse}
\subsubsection{Specimen characteristics}\label{sec-specimen-characteristics}

This section demonstrates the relationship between CVI processing and flexural strength, in which a pore length of $L=0.075$ mm is used. Based on the convergence testing in Sec. \ref{sec-convergence}, all tests in this section utilize a time step of $\Delta t_{\text{CVI}} = 0.1$ s, a CVI grid spacing of $\Delta z_{\text{CVI}} = L/(N-1)$, where $N=80$ is the number of grid points, and an FE mesh spacing of $\Delta z_{\text{FPB}} = L/N_e$, where $N_e=28$ is the number of elements in the vertical direction. This combination is chosen because of its ability to rapidly calculate solutions while yielding minimal inaccuracy (Fig. \ref{fig-spatialLsmall}, Table \ref{table-hcases}).

\begin{figure}
\centering
\begin{subfigure}{\textwidth}
  \centering
  \includegraphics[width=\linewidth]{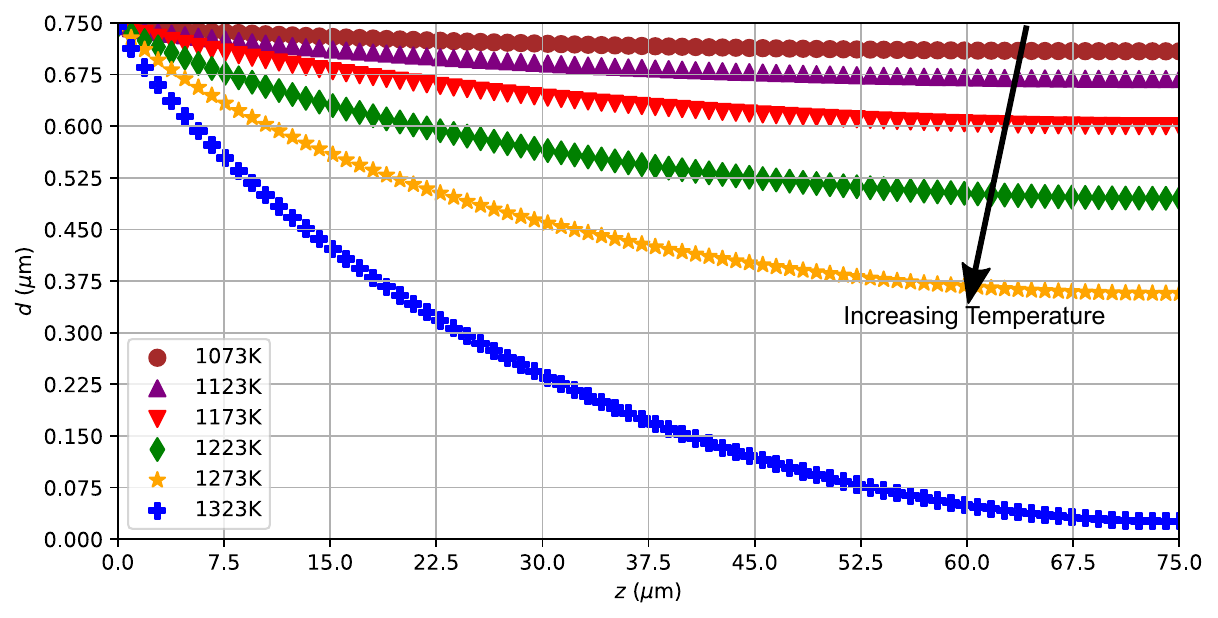}
  \caption{\doublespacing \ }
  \label{CVI-temp}
\end{subfigure}

\begin{subfigure}{\textwidth}
  \centering
  \includegraphics[width=\linewidth]{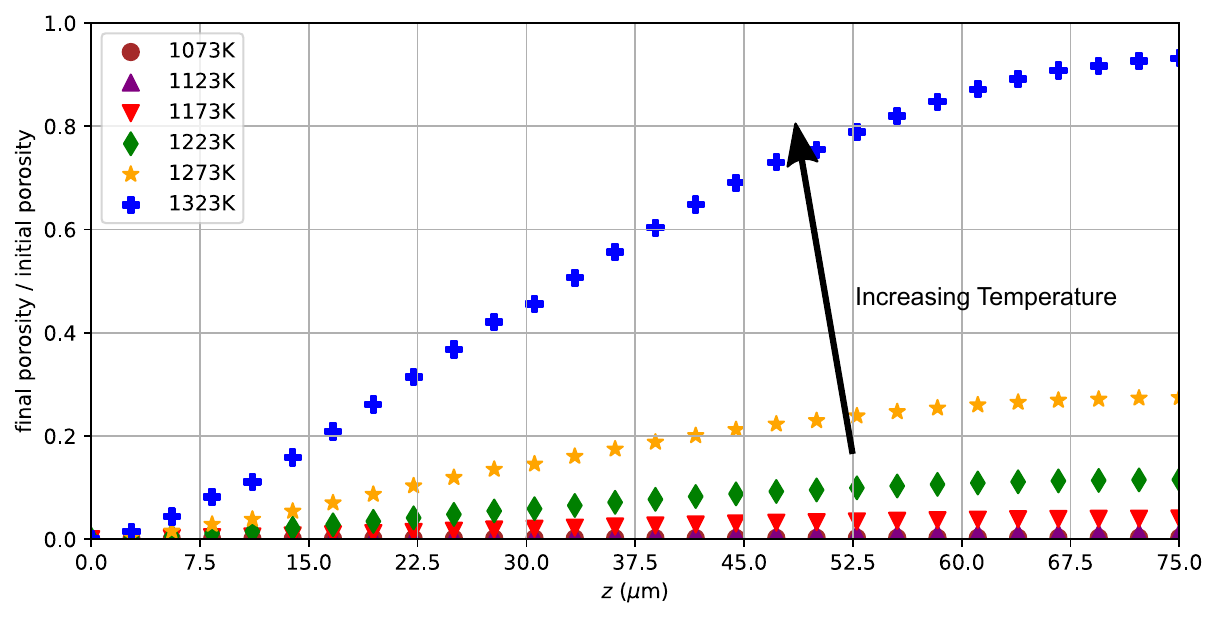}
  \caption{\doublespacing \ }
  \label{CVI-damage1}
\end{subfigure}
\caption{\doublespacing (a): Influence of temperature on infiltration depth. (b): Ratio between final porosity and initial porosity for various temperatures.}
\label{CVI-damage}
\end{figure}

The porosity information can be integrated with a four-way coupled thermo-mechanical damage model for a more complex, physics-based problem. 
$SiC$ based ceramics have initial porosities ranging from 9\% to 95\% resulting from different fabrication processes~\cite{Eom2013}. The phase field modeling of fracture can take any given porosity distribution as the initial condition of damage order parameter. The initial porosity for the entire specimen (assuming uniformly distributed pores) is defined as $r_{p0}^2/R_V^2$ where $r_{p0}$ is the initial pore radius and $R_V$ is the radius of the representative volume element. Therefore the radius of the RVE can be determined as
\begin{equation}
    R_V = \frac{r_{p0}}{\sqrt{\text{initial porosity}}}.
\end{equation}
Particularly, the cases considered in this study have initial porosities of 9\%, 30\%, 60\%, and 95\% as a representation of the wide range of $SiC$-based ceramic materials. The initial pore radius is $r_{p0}=\Phi_0/2 = 0.75 \ \mu $m. The relationship between CVI temperature and the final porosity profile of the specimen is illustrated in Fig.~\ref{CVI-damage}. At higher temperatures, the time needed for the surface of the pores to close decreases. Consequentially, the pore diameter farther into the depth remains relatively large. Note that the final porosity will vary as a function of the depth in the pore and is related the square of the final pore radius at a particular depth. This explains the widening of the gap between cases 1273 K and 1323 K when infiltration depth is converted to porosity. The difference between the infiltration depths of 1273 K and 1323 K is 0.93 times the difference between 1073 K and 1223 K. However, the difference between the final porosities of 1273 K and 1323 K is 2.55 times the difference between 1073 K and 1273 K.

\subsubsection{Four point bending test}

The weakening of material properties due to porosity, as well as the increase in damage across time, is represented by the degradation function Eq.~\eqref{Eq12}. The previous sections explored the relationships between various processing inputs and the resulting pore structure; this section describes the relationship between the porous structure and the resulting performance in a flexure test. According to the procedure described in Sec. \ref{properties-performancesection}, the ASTM C1161-18 four-point bending test~\cite{C282013} is simulated.  Each test is normalized against the fracture behavior of a specimen with no porosity ($\sigma_{f0} = 374.33$ MPa, $t_{\sigma_f0} = 4.97$ s).
\begin{table}[ht]
\centering
\begin{tabular}{llllll}
$T_{\text{CVI}}$ (K) & Solve time (s, hrs) & $t_f$ (s, hrs)
\\ \hline
$1073$ & 7433.47, 2.06& 3513.8, 0.98 \\
$1123$ & 1921.52, 0.53 & 839.5, 0.23 \\
$1173$ & 584.28, 0.16 & 227.0, 0.06 \\
$1223$ & 130.93, 0.04 & 68.5, 0.02 \\
$1273$ & 43.54, 0.01 & 22.8, 0.01 \\
$1323$ & 18.53, 0.01 & 8.2, 0.00
\end{tabular}
\qquad
\begin{tabular}{llllll}
Init. por. & $T_{\text{CVI}}$ (K) & $\sigma_f/\sigma_{f0}$ & $t_{\sigma_f}/t_{\sigma_f0}$
\\ \hline
$5\%$ & $1073$  & 0.999  & 1.000 \\
 & $1273$  & 0.999  & 1.002 \\
 & $1323$  & 0.999  & 1.004 \\
 \hline
$30\%$ & $1073$  & 0.999  & 1.000 \\
 & $1273$  & 0.999  & 1.004 \\
 & $1323$  & 0.912  & 1.153 \\
 \hline
$60\%$ & $1073$  & 0.999  & 1.000 \\
 & $1273$  & 0.984  & 1.062 \\
 & $1323$  & 0.565  & 1.525 \\
 \hline
$95\%$ & $1073$  & 0.999  & 1.000 \\
 & $1273$  & 0.920  & 1.137 \\
 & $1323$  & 0.001  & 0.004
\end{tabular}

\caption{\doublespacing CVI solve time, time to infiltration ($t_f$), normalized flexural strength $(\sigma_f/\sigma_{f0})$, and normalized time to fracture ($t_{\sigma_f}/t_{\sigma_f0}$) based on CVI temperature ($T_{\text{CVI}}$) and the specimen's initial porosity.}
\label{tab:cvi-fs}
\end{table}

\begin{figure}[ht]
    \centering
    \includegraphics[width=\linewidth]{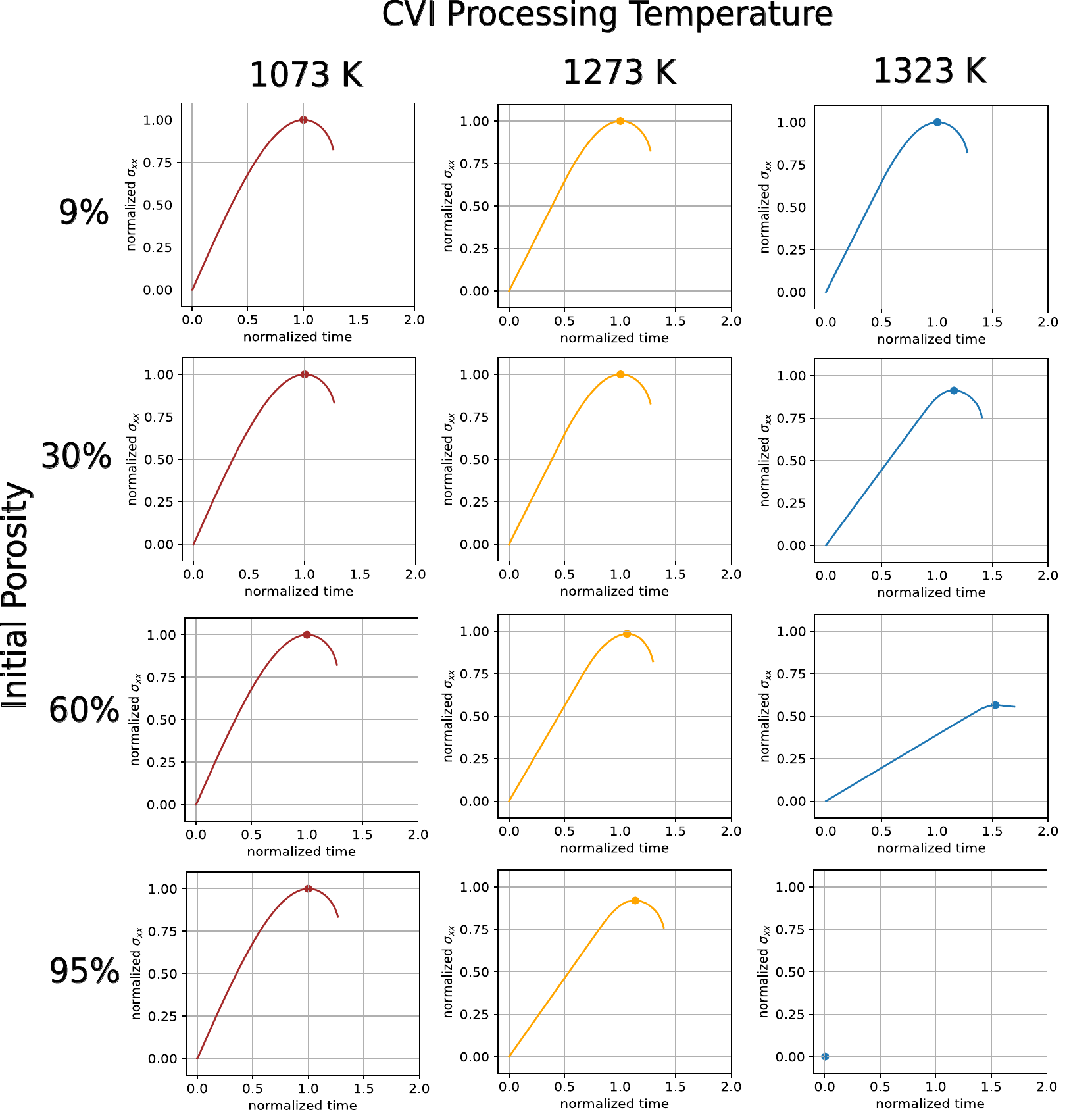}
    
    \caption{\doublespacing Time-dependent stress of $\alpha$-$SiC$ specimens with initial porosities 9\%, 30\%, 60\%, 95\% and CVI processing temperatures 1073 K, 1273 K, and 1323 K. \color{black}}
    \label{fig-porFS3}
\end{figure}

\color{black}Fig. \ref{fig-porFS3} is an illustration of the four point bending test results written in Table \ref{tab:cvi-fs}. \color{red}While there is no closed form for the relationships between these parameters, qualitative inferences can be drawn from the data. The fundamental tradeoff in a manufacturing environment is between infiltration time and flexural strength. Infiltration time is improved by increasing the CVI temperature, which improves the speed of the deposition reaction. However, the infiltration time cannot be improved ad libitum without compromising the resulting material's flexure performance. Except in cases where the specimen's porosity prior to CVI treatment is approximately less than 30\%, increasing the temperature past 1273 K leads to a crucial downgrade in flexural strength. This is because the initial condition for the damage parameter grows in proportion to the square of the pore radius. When the infiltration depth is low, small differences in the infiltration depth lead to large differences in the initial damage. On the other hand, when the infiltration depth is high, small differences in the infiltration depth lead to negligible differences in initial damage. As the processing temperature exceeds $1273$ K, the infiltration depth is less than half of the initial pore radius. Therefore, as temperature increases beyond this point, the initial damage increases significantly. \color{black}

Sensitivity to damage is likewise experienced by specimens with initial porosities exceeding 30\%. 
For instance, the flexural strengths with the temperatures $1073\text{ K} \text{ and }1323\text{ K}$ for a specimen with initial porosity 9\% are virtually identical. Therefore, for specimens of this type, a more relevant factor in considering the optimal manufacturing choice is CVI time of infiltration. On the other hand, the flexural strengths with the same processing parameters for specimens with an initial porosity of 60\% are $374.33\text{ MPa and } 221.40\text{ MPa}$ respectively. The specimen subjected to the latter CVI conditions fails under less stress when the initial porosity of the material prior to infiltration is high at 60\%. Therefore, optimizing time of infiltration should be sacrificed in favor of a more uniform densification for highly macroporous specimens.

\FloatBarrier
\section{Conclusion}\label{Sec4}
In this study, to account for the variability in strength associated with chemical vapor infiltration on $\alpha$-$SiC$ ceramics, the full hierarchical process of ICME has been utilized.
The influence of the chemical vapor infiltration process on the mesoscale pores, and its effect on macroscale flexural strength, is crucial information in a limited, decision-based manufacturing environment. The results of this study show that if the specimen being densified through CVI has a low initial porosity ($\approx 9\%$), the resulting flexural strength is relatively independent of the processing parameters. Therefore, the usage of a higher temperature and pressure is beneficial to reduce the overall time to infiltration. Particularly, for a specimen with an initial porosity of $9\%$, an increase in the CVI temperature from 1073 K to 1323 K leads to an increase in initial damage from 0.02\% to  8.34\%. This is associated with an unidentifiable decrease in flexural strength. However, a higher initial porosity ($\gtrapprox 30\%$) will increasingly have a flexure response that is in closer relationship to the processing parameters. Particularly, for a specimen with an initial porosity of $60\%$, an increase in the CVI temperature from 1073 K to 1323 K leads to an increase in initial damage from 0.16\% to 55.58\%. Consequentially, this temperature change leads to a decrease in flexure strength from 374.33 MPa to 221.40 MPa, which is a significant 43.5\% decrease. In this case, the quality of densification competes with the quality of infiltration time, and lower processing temperatures should have higher priority in the context of a manufacturing decision. These results are self-consistent, and the formulation used to collect them is valid in relationship to the surrounding literature. Therefore, the ICME framework offered in this study is a reliable and rapid tool in evaluating the mechanical performance of diverse kinds of chemical vapor infiltrated $SiC$-based ceramic materials.

\section{Acknowledgements}\label{Sec5}
The authors are very grateful and wish to acknowledge David Hicks, Amberlee Haselhuhn, Anthony Koumpias, and Bradley Friend from LIFT for their ongoing support and helpful conversations. Funding for this project was provided, in part, by LIFT, the Detroit-based national manufacturing innovation institute operated by ALMMII, the American Lightweight Materials Manufacturing Innovation Institute, a Michigan-based nonprofit, 501(c)3 as part of ONR Grant N00014-21-1-2660.

\bibliography{bib}
\bibliographystyle{ieeetr}

\end{document}